\documentclass[letterpaper]{JHEP3}
\usepackage{epsfig}
\usepackage{graphicx}
\usepackage{amssymb,amsmath}
\usepackage{epstopdf}

\def\be{\begin{eqnarray}}
\def\ee{\end{eqnarray}}

\title{Magnetic properties of dense Holographic QCD}
\author{Oren Bergman\thanks{On sabbatical leave from the Department of Physics,
Technion, Haifa 32000, Israel.}\\
School of Natural Sciences\\
Institute for Advanced Study\\
Princeton, NJ 08540, USA\\
\email{bergman@sns.ias.edu,bergman@physics.technion.ac.il}}

\author{Gilad Lifschytz\\
Department of Mathematics and Physics and CCMSC \\
University of Haifa at Oranim\\
Tivon 36006, Israel \\
\email{giladl@research.haifa.ac.il}}

\author{Matthew Lippert\\
Department of Physics\\
Technion, Haifa 32000, Israel\\
{\rm and}\\
Department of Mathematics and Physics \\
University of Haifa at Oranim\\
Tivon 36006, Israel\\
\email{matthewslippert@gmail.com}}

\date{}

\abstract{We investigate the Sakai-Sugimoto model at nonzero baryon
chemical potential in a background magnetic field both in the confined phase
and in the deconfined phase with restored chiral symmetry.
In this case the 8-brane Chern-Simons term becomes important.
In the confined phase it generates a gradient of the pseudo-scalar ``pion",
which carries a non-vanishing baryon charge.
Above a critical value of the chemical potential there
is a second order phase transition to a mixed phase which includes 
also ordinary baryonic matter.
However, at fixed baryon charge density the matter is purely ``pion"-gradient above 
a critical magnetic field.
In the deconfined chiral-symmetric phase at nonzero chemical potential
the magnetic field induces an axial current.
We also compute the magnetization of the baryonic matter 
and find that it is paramagnetic in all three phases but with nonlinear
behavior at large magnetic field.}

\begin{document}

\section{Introduction}

The behavior of QCD under external conditions is an interesting and
physically relevant problem. 
For example, at high temperature the ground state is believed to be
a deconfined quark-gluon plasma, and at high density it is believed 
to be a color-superconductor. The former is relevant for the understanding
of the physics at RHIC, and the latter may be relevant for the physics 
of dense stellar objects such as neutron stars.
Background electromagnetic fields provide another kind of external condition.
Their effect on the QCD ground state, in particular in the
charged flavor sector, is possibly relevant for magnetars, which are neutron 
stars with very large magnetic fields.
However, it is often the case that in the physically relevant regimes QCD
is strongly coupled, and we cannot reliably use perturbation 
theory.\footnote{This appears to be the case at RHIC, for example.}
Lattice gauge theory has been very successful for studying equilibrium
properties of QCD at nonzero temperature, but it doesn't do so well
at nonzero chemical potential, and it is not equipped to handle 
real-time transport properties, such as those expected in a background electric field. Another possibility which is non-perturbative is to use some 
truncation of the Schwinger-Dyson equations to study QCD.
An alternative approach to strongly-coupled gauge theories has
emerged in recent years from string theory via the AdS/CFT
correspondence and its generalization to gauge/gravity holographic duality.
The holographic approach is well-poised to address questions related to
external conditions since these translate simply to boundary conditions on
internal fields in the bulk. 
For example, the temperature corresponds to the size of the Euclidean time
dimension at the boundary, and a chemical potential associated with
a conserved current corresponds to the boundary value of the bulk gauge field
dual to the current.
Determining how an external condition affects
the ground state of the field theory is then just a matter of solving the bulk
equations of motion with the appropriate boundary conditions.

Due to asymptotic freedom, the holographic dual of QCD cannot be
a (super)gravity theory alone and must include all the closed string excitations.
Neverthless, it may be useful to study gravitational models as holographic
models of low-energy properties of large $N_c$ QCD, in the hope of 
eventually embedding them in string theory with the correct UV properties.
This has been the approach of the so-called ``bottom-up" models.
The ``top-down" approach, in contrast, is to consider the full string theory
in a background where the low-lying excitations resemble those of QCD
and in which one can consistently study the supergravity limit.
This does not give QCD, but these kinds of models share many of its 
low-energy properties. It is worth emphasizing again that we are studying these models in the hope that they resemble QCD, but the analysis is of these models only. 
The closest so far to QCD is the Sakai-Sugimoto model \cite{SS}.
This model consists of $N_c$ 4-branes wrapping a circle with anti-periodic
boundary conditions for the fermions, $N_f$ 8-branes at a point on the circle,
and $N_f$ anti-8-branes at another point on the circle. 
The low-lying open string excitations are precisely those of $SU(N_c)$ Yang-Mills
theory with $N_f$ flavors of massless quarks.
The holographic limit corresponds to $N_c\rightarrow\infty$, and $N_f$
is kept finite so that the 8-branes are treated as probes in the near-horizon
background of the 4-branes. In this limit the background is capped off in the IR,
which corresponds to confinement in the gauge theory, and the 8-branes
and anti-8-branes connect into U-shaped 8-branes, which corresponds
to chiral symmetry breaking in the gauge theory.
This model has also been studied in various external conditions, including
nonzero temperature \cite{ASY}, 
nonzero baryon chemical potential \cite{BLL1,finite_density}, 
and background electric and magnetic fields \cite{BLL2,Johnson_Kundu,Zahed2}, 
in which it exhibits many properties that are expected of QCD.

Background magnetic fields are particularly interesting in that they may
be physically relevant in neutron stars, where they can reach values of about
$10^{15}$ Gauss. On the theory side, background magnetic fields have
some interesting effects on the QCD ground state.
One effect is the catalysis of chiral symmetry breaking by a strong magnetic field
\cite{Miransky}. The basic mechanism for this is that in a strong magnetic field
all the quarks sit in the lowest Landau level, and the dynamics are effectively
1+1 dimensional. 
The effect of the magnetic field on the quark condensate was studied in 
\cite{Shushpanov:1997sf}.
The effect of a background magnetic field in the
Sakai-Sugimoto model was studied in \cite{BLL2, Johnson_Kundu},
where the catalysis of chiral symmetry breaking was demonstrated explicitly.
In particular it was shown that the critical temperature for the restoration
of chiral symmetry increases with the magnetic field and approaches
a finite temperature at infinite magnetic field.
%\footnote{In the D3-D7 model the critical temperature diverges  
%at a finite magnetic field \cite{Erdmenger}.}

In this paper we will be interested in the effects of a background magnetic field
at nonzero baryon chemical potential. 
This question was recently studied in the the low-energy effective field theory 
\cite{Son_Zhitnitsky, axial_current, Son_Stephanov},
where it was shown that the triangle anomaly leads to interesting 
effects.\footnote{For other interesting anomaly-induced effects at finite density and 
magnetic field see \cite{Harvey}).}
In the deconfined chiral-symmetric phase the combination of a magnetic
field $B$ and a nonzero baryon chemical potential $\mu_B$ leads to 
a non-zero axial current density \cite{axial_current}
\be
\label{axial_current_son}
j_A = {e\over 2\pi^2} \mu_B B \,.
\ee
This current is generated purely by fermionic zero modes and is therefore topological
in nature. The result is therefore also exact.
In the confined phase the anomaly leads to a non-trivial
pion gradient and an associated baryon charge density \cite{Son_Stephanov},
\be
\label{pion_gradient_son}
\nabla\pi^0 = {e\over 4\pi^2 f_\pi} \mu_B B \;\; , \;\;
d = {e\over 4\pi^2 f_\pi} B \cdot \nabla\pi^0 \,.
\ee

We will show that similar effects occur in the one-flavor Sakai-Sugimoto model.
In this model the anomaly is encoded in the five-dimensional
Chern-Simons term of the 8-brane action. The model does not include a true
electromagnetic field, but we can mimic the effect of a non-dynamical
(background) electromagnetic field using the non-normalizable mode
of the 8-brane worldvolume gauge field. This field is actually dual to the
baryon current in the gauge theory but is equal in the one-flavor case to the electric current.
We will therefore use the same bulk gauge field,
but different components, to describe both the baryon chemical potential
and the background magnetic field.
We will show that these source a third, normalizable component of the gauge field via the
Chern-Simons term.
In the low-temperature confining background this field has a nonzero
boundary value, which is interpreted as the gradient of the $U(1)_A$ pseudo-scalar
meson, {\em i.e.} the $\eta'$.
This also leads to a baryon number charge density.
For small magnetic fields, our result agrees with (\ref{pion_gradient_son}) 
adapted to the $U(1)_A$ sector.
Furthermore, we will show that there is a phase transition at a critical value of the chemical
potential to a mixed phase of ordinary baryonic matter and pseudo-scalar gradient matter.
In the mixed phase the relative proportion of ordinary baryonic matter at fixed chemical potential
decreases with the magnetic field.
In the high-temperature deconfining background, in the restored chiral-symmetry phase,
the induced gauge field has a vanishing boundary value, and the leading 
asymptotic behavior corresponds to an axial current density,
which agrees with (\ref{axial_current_son}).

We will make two simplifying assumptions about the 8-brane embedding,
which do not affect our results qualitatively.
First, we will consider only the one-flavor case, in which the 8-brane worldvolume
gauge field is abelian. 
This will allow us to use the full DBI action for the 8-brane.
Second, we will consider only the antipodal 8-brane embedding.
This means that in the low-temperature confining background
the tip of the 8-brane coincides with the tip of the space.
This will also simplify the analysis with sources since the embedding will
remain smooth. This should not affect the qualitative results since this embbeding is smoothly connected to the non-antipodal embedding.\footnote{in fact there is a scaling argument connecting all of these embbedings, see the second reference in \cite{finite_density}.}
In the high-temperature deconfining background it implies that the preferred
embedding is the disconnected 8-brane-anti-8-brane configuration, in
which chiral symmetry is restored.
Finally, to avoid clutter we will work mostly with dimensionless quantities by absorbing
appropriate factors of $R$ and $\alpha'$.
We will denote such quantities by lower case letters.
For example the dimensionless coordinates are $u=U/R$ and $x_\mu=X_\mu/R$.
A translation table between the dimensionless and physical versions of
the relevant quantities is provided in the appendix, but we will also 
give the translation whenever a new quantity is introduced.

The rest of the paper is organized as follows:
In section 2 we review the relevant features of the Sakai-Sugimoto model 
at nonzero baryon chemical potential.  In section 3 we analyze the magnetic
properties of the confined phase, including the pseudo-scalar gradient and baryon 
charge density, as well as the magnetization. In section 4 we study the magnetic
properties of the deconfined phase. Section 5 contains our conclusions.

\medskip

\noindent {\underline{\em Note:}} 
While our paper was being completed the paper
\cite{Son_new}, with which there is some overlap, came to our attention. 

\section{Review of finite density HQCD}

\subsection{Confined phase}

The Euclidean background dual to the confining phase is given by
\be
\label{SS_background}
ds^2 &=& u^{3/2} \left((dx_0^E)^2 + d{\bf x}^2
+ f(u) dx_4^2\right)
+ u^{-3/2}\left({du^2\over f(u)}
+ u^2 d\Omega_4^2\right) \nonumber\\
e^{\Phi} &=& g_s u^{3/4} \; , \;
F_4 = 3\pi (\alpha')^{3/2}  N_c \, d\Omega_4 \,,
\ee
where $x_4$ is a compact coordinate with periodicity $2\pi R_4$, and 
\be
f(u) = 1- {u_{KK}^3\over u^3} \; , \; u_{KK} = {4R^2\over 9R_4^2} \,.
\ee
The curvature radius of the space is given by
\be
R = (\pi g_s N_c)^{1/3} \sqrt{\alpha'} \; ,
\ee
and this is related to the four-dimensional 't Hooft coupling 
\be
\lambda = {4\pi g_s N_c \sqrt{\alpha'}\over R_4}\,.
\ee
The antipodal embedding of the 8-brane in this background has a U shape
that satisfies $x_4'(u)=0$, with the tip at $u_{KK}$.
Other than the embedding scalar field $x_4$, the 8-brane worldvolume theory
contains fermions and a gauge field. We will ignore the fermions.
The gauge field has, in general, both a vector and an axial part depending on the parity under 
exchanging the two halves of the embedding,
\be
a_M(x^\mu,u) = a_M^V(x^\mu,u) + a_M^A(x^\mu,u) \,.
\ee
The physical gauge field is $A_M=a_M R/(2\pi\alpha')$.
We will consider only fields that are uniform on the $S^4$, so the index
$\mu$ runs over $0-3$.
There is a discrete spectrum of normalizable radial modes corresponding 
to various low-spin mesons.
In particular the zero mode of $a_u^A$ is identified with the massless pseudo-scalar
corresponding to the Goldstone boson of the broken chiral symmetry.
For a single flavor this is the $\eta'$.\footnote{The anomalous mass of the $\eta'$
is suppressed at large $N_c$.}
However there is some freedom in identifying the mesons due to the gauge symmetry.
A particularly nice gauge choice, that preserves the four-dimensional Lorentz
symmetry, is $a_u=0$ \cite{SS}.
In this gauge the pseudo-scalar reappears in the zero mode of $a^A_\mu$,
\be
a_\mu^A(x^\mu,u) = \partial_\mu \varphi(x^\mu) \psi_0(u) + 
\mbox{higher modes} \,,
\ee
where
\be
\psi_0(u) = {2\over\pi}\, \arctan\sqrt{{u^3\over u_{KK}^3}-1}\,.
\ee
The physical pseudo-scalar field is $\eta'(X) = f_\pi \varphi(x) R^2/(2\pi\alpha')$,
where $f_\pi$ is given by\footnote{The pion decay constant was determined
in terms of the parameters of the model by comparing the {\em non-abelian}
8-brane Yang-Mills action with the standard {\em non-linear} sigma model.} 
\be
f_\pi^2 = {N_c u_{KK}^{3/2}\over 8\pi^4\alpha'} \,.
\ee
Note that the axial zero mode has a normalizable field strength.
By contrast, the zero mode of the vector part of the gauge field $a_\mu^V(x^\mu)$ is 
$u$-independent and therefore non-normalizable.
In general, this corresponds to a source for the vector (baryon) current in the 
boundary gauge theory.
In particular, the asymptotic value of the $x^\mu$-independent part of
$a_0^V$ is identified with the baryon chemical potential,
\be 
a_0^V(u\rightarrow \infty) = \mu \,.
\ee
In our convention the baryon charge of a quark is 1, rather than $1/N_c$.

For a static and uniform baryon charge distribution 
the (Euclidean) 8-brane DBI action per unit 4-volume of spactime 
is given by
\be 
S_{DBI} =   {\cal{N}} \int_{u_{KK}}^\infty du \, u^{5/2} \sqrt{ \frac{1}{f(u)}
-(a_0^{V\prime}(u))^2} \,,
\ee
where the overall normalization is given by
\be
\label{normalization}
{\cal N} = 2 \Omega_4 T_{D8} R^5  = {N_c\over 6\pi^2}\, {R^2\over (2\pi\alpha')^3} \,.
\ee
The factor of 2 corresponds to the two halves of the embedding.
The resulting equation of motion for the gauge field is given by
\be
{d\over du}\left[{u^{5/2}
\sqrt{f(u)} a_0^{V\prime}(u)\over\sqrt{1 - f(u)(a_0^{V\prime}(u))^2}}\right] = 0 \,.
\ee
Integrating once gives
\be
\label{a0prime_confined}
a_0^{V\prime}(u) = {1\over \sqrt{f(u)}} {d\over\sqrt{u^5 + d^2}} \,,
\ee
where $d$ is the constant of integration.
The asymptotic solution at large $u$ is then
\be
\label{smallz}
a_0^V(u) \approx 
\mu - {2\over 3} {d\over u^{3/2}} \,.
\ee
Since the action (per unit 4-volume) of the solution defines the grand potential
(per unit 3-volume) of the gauge theory at a fixed $\mu$, 
we identify the constant $d$ as the baryon charge density.
The physical chemical potential is $\mu_B = \mu R/(2\pi\alpha')$,
and the physical baryon charge density is
$D= d (2\pi\alpha'{\cal N}/R)$.

In the absence of sources the only solution is a constant
\be
\label{vac_solution}
a_0^V(u) = \mu \,.
\ee
In this case the gauge field is pure gauge, and therefore the physics does
not depend on the value of $\mu$.
A second solution becomes possible when one includes 
sources at the tip corresponding to 4-branes wrapped on the $S^4$.
These 4-branes are precisely the baryons of the model.
A single baryon carries $N_c$ units of baryon charge.
Assuming a uniform distribution of 4-branes with positive
number density $n_{4}$,
and assuming that the 4-branes are well-separated in space so that we
can ignore interactions between them, the source action per unit 4-volume is given by
\be
S_{D4} = {\cal N} \left(n_{4} m_{4}
- n_{4} N_c \int du\, a_0^V(u) \delta(u-u_{KK}) \right)\,.
\ee
The first term comes from the 4-brane DBI action, where
$m_4$ is the mass of a wrapped 4-brane, {\em i.e.} a baryon, located
at $u=u_{KK}$,
\be
m_4 = {1\over 3} N_c u_{KK} \,.
\ee
The physical 4-brane mass and density are given by $M_4 = m_4 R/(2\pi\alpha')$
and $N_4 = n_4(2\pi\alpha'{\cal N}/R)$.
The second term in the source action comes from the $N_c$ strings that connect each
4-brane to the 8-brane (or equivalently from the 8-brane CS term, if we describe 
the 4-branes as instantons in the 8-branes \cite{baryons}).
This relates the baryon charge density to the 4-brane number density as
\be
\label{sources}
d = N_c n_{4} \,.
\ee
We can then determine this number for the solution by
extremizing the action with respect to $n_{4}$. This gives a condition
on the gauge field at the tip,\footnote{This can also be seen by requiring a 
consistent interpretation of the themodynamic potentials \cite{BLL1}}
\be
\label{tip_condition}
a^V_0(u_{KK}) = {m_{4}\over N_c} \,,
\ee
which implies that the solution exists only for 
$\mu > m_{4}/N_c$.\footnote{For anti-four-branes this is 
$-m_{4}/N_c$.}
This has the obvious phenomenological interpretation that, at low temperature,
baryons can only appear when the chemical potential is high enough
to produce them.
Furthermore, it is easy to see from the form of 
the action that this ``nuclear matter" solution dominates over the 
vacuum solution (\ref{vac_solution}).
In other words, nuclear matter forms as soon as it can, and there is a phase
transition at $\mu_c = m_{4}/N_c$. 
The relation between the baryon charge density and chemical potential
is obtained by integrating (\ref{a0prime_confined}),
\be
\mu = \mu_c +  \int_{u_{KK}}^\infty {du\over \sqrt{f(u)}}{d\over \sqrt{u^5+d^2}}\,.
\ee
Near the critical point we get a linear relation
\be
d \approx {3 u_{KK}^{3/2}\over \pi}(\mu - \mu_c) \,,
\ee
which implies that the phase transition is marginally second-order.
This is different from the 
expected first-order transition in QCD.
However the result is reasonable since we have ignored baryon interactions.

\subsection{Deconfined phase}

The Euclidean background for the deconfined phase is given by
\be
\label{deconfined_background}
ds^2 = u^{3\over 2} \left(f(u)(dx_0^E)^2 + d{\bf x}^2
+ dx_4^2\right)
+ u^{-{3\over 2}}\left({du^2\over f(u)}
+ u^2 d\Omega_4^2\right)\,,
\ee
with the same dilaton and RR field as before, and with
\be
f(u) = 1 - {u_T^3\over u^3} \,,
\ee
where $u_T$ is related to the temperature, {\em i.e.} the inverse periodicity
of the Euclidean time $x_0^E$, as $u_T=(4\pi/3)^2 R^2 T^2$.
For $T>1/(2\pi R_4)$ this phase dominates over the confined phase
and the the theory deconfines.
In this phase the 8-branes in general have two possible embeddings:
a U-shaped embedding similar to the one in the confined phase, and a parallel
8-brane-anti-8-brane embedding \cite{ASY}.
We will consider only the parallel embedding, which is the dominant phase
at all temperatures in the antipodal case.\footnote{Below a certain value of 
the asymptotic 8-brane-anti-8-brane separation there is a range 
of temperatures for which U-shaped embedding dominates, and the theory
realizes an interesting intermediate phase of deconfinement and chiral symmetry
breaking. We will not consider this phase here.}
In this embedding there are two independent gauge fields $a_\mu$
and $\bar{a}_\mu$, associated with the 8-brane and anti-8-brane, respectively.
As before we work in the gauge $a_u=\bar{a}_u=0$.

There are a number differences from the confined phase.
First, the spectrum of normalizable solutions is not discrete, so
there is no particle (meson) interpretation \cite{PSZ}.
Second, both zero modes are non-normalizable and therefore correspond
to two sets of parameters in the gauge theory.
We will set the axial parameters to zero.
Our boundary conditions at infinity are therefore
\be
a_0(\infty) = \bar{a}_0(\infty) = \mu \,.
\ee
Another important difference is that for regularity at the horizon we must impose 
the boundary conditions
\be
a_0(u_T) = \bar{a}_0(u_T) = 0 \,.
\ee
The two gauge fields are therefore equal, and the total action for the 8-brane
and anti-8-brane is given by
\be 
S_{DBI} =   {\cal{N}} \int_{u_T}^\infty du \, u^{5/2} \sqrt{1-(a_0'(u))^2} \,,
\ee
where the normalization is the same as in (\ref{normalization}), with the factor of
2 accounting for the two branes.
The solution now satisfies\footnote{This can be expressed in terms of a 
hypergeometric function \cite{PS}.}
\be
\mu = a_0(\infty) = \int_{u_T}^\infty du {d\over\sqrt{u^5 + d^2}} \,.
\ee
In this phase matter is made up of deconfined quarks and
begins to form immediately at nonzero $\mu$.
For small $\mu$ the density is given by
\be
d \approx {3 u_T^{3/2}\over 2} \mu \sim T^3 \mu \,.
\ee

%%%%%%%%%%%%%%%%%%%%%%%%%%%%%%%%%%%%%%%%%%%%%

\section{Magnetic properties of the confined phase}

To mimic the effect of a background magnetic field we turn on
a background value for the zero mode of a spatial component of the vector gauge field,
\be
a_3^V(x_2,u) = hx_2 \,.
\ee
The physical magnetic field is $H = h/(2\pi\alpha')$.   
Since $a_0^V(u)\neq 0$, the five-dimensional CS term, which comes
from the 8-brane CS coupling to $F_4$, will source the axial field $a_1^A(u)$.
As we saw in the previous section, the boundary value of this field
corresponds to the (constant) gradient of the pseudo-scalar 
(in the $x^1$ direction in this case),
\be
a_1^A(\infty) = \nabla\varphi \,.
\ee
This corresponds to a field, rather than an external parameter, in the gauge theory,
since the zero mode of the axial field is normalizable.
Its value is therefore determined by extremizing the action.
Furthermore, since $a_1^A$ is an axial field, it must vanish at the tip,
\be
a_1^A(u_{KK}) = 0 \,.
\ee
The action with all the relevant fields is $S_{DBI}+S_{CS}$, where\footnote{Our ansatz is that $a_0(u)$ and $a_1(u)$ depend only on $u$.  However, we retain two terms in the CS action with derivatives with respect to $x_2$, since they contribute to the equations of motion.}
\be 
S_{DBI} &=&   {\cal{N}} \int_{u_{KK}}^\infty du \, u^{5/2} \sqrt{\left(\frac{1}{f(u)}
-(a_0^{V\prime}(u))^2 +(a_1^{A\prime}(u))^2 \right)\left(1+{h^2\over u^3}\right)}\\
S_{CS} &=&-{\cal N}
\int_{u_{KK}}^\infty du\left(\partial_2a_3^V a_0^V(u) a_1^{A\prime}(u) 
-\partial_2a_3^V a_0^{V\prime}(u) a_1^A(u)-a_{3}^{V}\partial_2a_0^Va_1^{A\prime}+a_{3}^{V}\partial_2a_1^{A}a_{0}^{V\prime}\right), \nonumber
\ee
The corresponding integrated equations of motion are given by
\be 
\label{EOMa0}
{\sqrt{u^5+ h^2 u^2}\, a_0^{V\prime}(u) 
\over\sqrt{{1\over f(u)} - (a_0^{V\prime}(u))^2 + (a_1^{A\prime}(u))^2}}
&=& 3ha_1^A(u) + N_c n_{4}\\
\label{EOMa1}
{\sqrt{u^5+ h^2 u^2}\, a_1^{A\prime}(u) 
\over\sqrt{{1\over f(u)} - (a_0^{V\prime}(u))^2 + (a_1^{A\prime}(u))^2}}
&=& 3ha_0^V(u) +c \,,
\ee
where the constant of integration in the $a_0^V$ equation has been identified
with the density of 4-brane sources as before, and the constant of integration in the 
$a_1^A$ equation will be determined shortly.

%%%%%%%%%%%%%%%%%%%%%%%%%%%%%%%%%%%%%%%%%%%%%%%%%%

% O: New paragraph about gauge invariance, and a footnote about agreement with the 5d current
The action as written above is a bit problematic however.
As is well known, in the presence of a boundary the CS action is not invariant under gauge transformations
that do not vanish at the boundary,
and requires the addition of a boundary action whose gauge transformation cancels 
that of the bulk action \cite{Elitzur:1989nr}. 
This will be true also in the infinite volume case if we allow for gauge transformations
that do not vanish at infinity, which we do.
In our case the gauge is partially fixed by $a_u=a_2=0$, and by the identification of 
$a_0^V(\infty)$ and $a_1^A(\infty)$ with $\mu$ and $\nabla\varphi$, respectively.
This leaves a residual gauge symmetry that depends only on $x_3$, under which
$a_3^V\rightarrow a_3^V + \partial_3\Lambda(x_3)$.
Under this transformation the CS action above transforms by (restoring the integration over the $x_\mu$,
and integrating by parts symmetrically):
\be
\delta S_{CS} =  {{\cal N}\over 2} \int dx_0 dx_1 dx_3 \, \partial_3 \Lambda(x_3)
\left[ \left.  \int du\, a_{[0}^V a_{1]}^{A\prime}\right|_{x_2\rightarrow -\infty}^{x_2\rightarrow \infty}
+ \left.\int dx_2 \, \partial_2 a_{[0}^V a_{1]}^A\right|_{u\rightarrow\infty}\right] \,.
\ee
In particular, the first term will be non-vanishing on shell since the gauge transformation does not depend on $x_2$.
Subsequently we need to add a boundary action at the boundaries $x_2\rightarrow \pm\infty$
and $u\rightarrow\infty$ that has the form
\be
S_{bndy} = -{{\cal N}\over 2} \left[\left. \int dx_0 dx_1 dx_3 du \, a_3^V a_{[0}^V a_{1]}^{A\prime}\right|_{x_2\rightarrow -\infty}^{x_2\rightarrow \infty}
+ \left. \int d^4x \, a_3^V \partial_2 a_{[0}^V a_{1]}^A\right|_{u\rightarrow\infty}\right] \,.
\ee
After integrating by parts the last two terms in $S_{CS}$, the 
sum of the CS and boundary actions becomes
\be
S_{CS} + S_{bndy} = -{\cal N} \int d^4x du \, \left[{3\over 2}\partial_2 a_3^V a_{[0}a_{1]}^{A\prime}
- {1\over 2} a_3^{V\prime}a_{[0}^V\partial_2 a_{1]}^A\right] \,,
\ee
which is manifestly invariant under the residual gauge 
transformation.\footnote{Note that with this modification the four dimensional current densities can
be expressed as integrals over the radial direction of the appropriatelly
defined five dimensional current densities. The five dimensional
current densities %have an ambiguity where one can add to it any constant times 
are defined in general only up to the addition of an arbitrary constant multiple of $\star(F\wedge F)$.
The restriction of the gauge symmetry above fixes this constant to be ${1\over 2}$.}

%%%%%%%%%%%%%%%%%%%%%%%%%%%%%%%%%%%%%%%%%%%%%%%%%%

The baryon charge and currents are defined by
\begin{equation}
J^{\mu}(x)=\frac{\delta S_{eom}}{\delta A_{\mu}(x,u=\infty)}
\end{equation}
Where $S_{eom}$ is the value of the action on the equation of motion.
This can be computed by 
\begin{equation}
\delta S_{eom}=\int \sum_{i}\frac{\delta {\cal L}}{\delta \partial_{i} A}\delta \partial_{i} A + \frac{\delta {\cal L}}{\delta  A}\delta A
\end{equation}   
By integrating by parts and using the equation of motion we find
%\begin{equation}
%\delta S_{eom} = \int \sum_{i=u,x_{2}}\partial_{i}\left(\frac{\delta {\cal L}}{\delta \partial_{i} A}\delta A\right)
%\end{equation}
%We use that $\delta A(x_{2}\rightarrow \pm \infty)\rightarrow 0$ fast enough.
%so we are left with
\begin{equation}
J^{\mu}(x)=\lim_{u \rightarrow \infty}\left(\frac{\delta {\cal L}}{\delta \partial_{u} A_{\mu}(x)}\right)
\end{equation}
where the right hand side is evaluated on the equation of motion.

We can now  read off the (dimensionless) baryon charge density 
\be
d = N_c n_{4} + \frac{3}{2} h a_1^A(\infty) = N_c n_{4} + \frac{3}{2} h\nabla\varphi \,.
\ee
The origin of the second term can be understood as an additional 4-brane charge 
inside the 8-brane, which is due to the orthogonal worldvolume 
field strengths in the $(x^2,x^3)$ and $(u,x^1)$ directions.
We can likewise get the (dimensionless) axial current density from (\ref{EOMa1}),
\be
j_A = c +\frac{3}{2} ha_0^V(\infty) = c + \frac{3}{2} h\mu \,.
\ee
Recall however that we still need to extremize the action with respect to
$\nabla\varphi$. This has the effect of setting $j_A=0$, and therefore $c=-\frac{3}{2} h\mu$.

We can simplify the equations of motion considerably as follows.
First, dividing (\ref{EOMa0}) by (\ref{EOMa1}) gives
an expression that can easily be integrated to give the relation
\be
\label{simplifying_relation}
\frac{3}{2} h(a_0^V(u))^2 -\frac{3}{2} h\mu a_0^V(u) = \frac{3}{2} h(a_1^A(u))^2 + N_c n_{D4} a_1^A(u) + \kappa \,,
\ee
where $\kappa$ is a constant that depends on the type of solution.
As in the zero-magnetic field case, there are two types of solutions,
with and without 4-brane sources.
In the sourced case there is an additional condition on $a^V_0$ at the tip given by
(\ref{tip_condition}).
Using the values at the boundary in the sourceless case, and at the tip
in the sourced case, we get
\be
\kappa = \left\{
\begin{array}{ll}
-\frac{3}{2}h(\nabla\varphi)^2 & \;\; \mbox{sourceless case} \\[5pt]
-\frac{3}{2} h {m_{4}\over N_c} \left(\mu - {m_{4}\over N_c}\right) & \;\; \mbox{sourced case} \,.
\end{array}\right.
\ee
Next, define a new coordinate
\be
y = \int_{u_{KK}}^u 
{3h d\tilde u\over\sqrt{f(\tilde u)}\sqrt{\tilde u^5\left(1+{h^2\over \tilde u^3}\right) + 
\left(N_c n_{4}\right)^2 - \left(\frac{3}{2}  h\mu\right)^2
-6h\kappa}} \,.
\ee
Using the relation (\ref{simplifying_relation}), and some algebra, 
the equations of motion then reduce to
\be
\label{simple_eqn_1}
a_0^{V\prime}(y) &=& a_1^A(y) + {N_c n_{4}\over 3h}  \\[5pt]
\label{simple_eqn_2}
a_1^{A\prime}(y) &=& a_0^V(y) - {\mu\over 2} \,,
\ee
where the derivative is with respect to $y$.
Let us now analyze the two types of solutions.

\subsection{Pseudo-scalar gradient phase}

\begin{figure}[htbp]
\begin{center}
\begin{tabular}{cc}
\epsfig{file=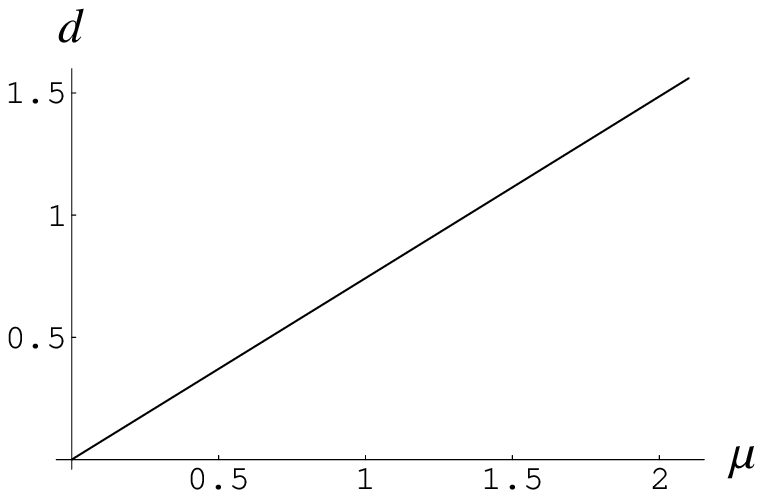,scale=0.8} \;\;\;\;\; &
\epsfig{file=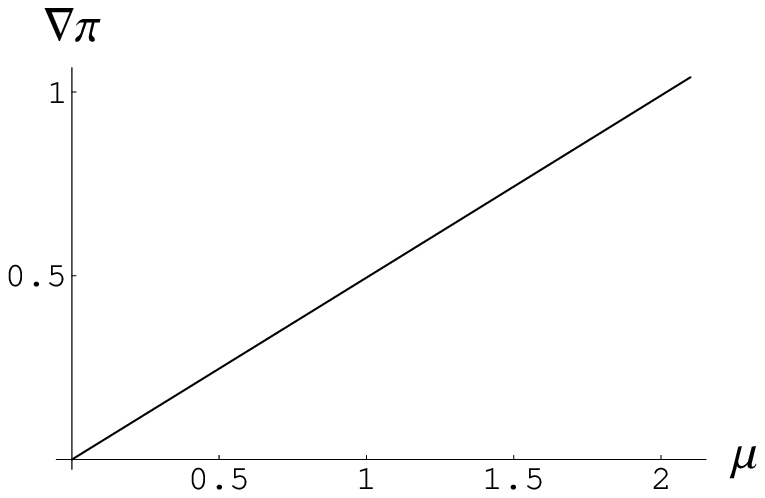,scale=0.8} \\
\epsfig{file=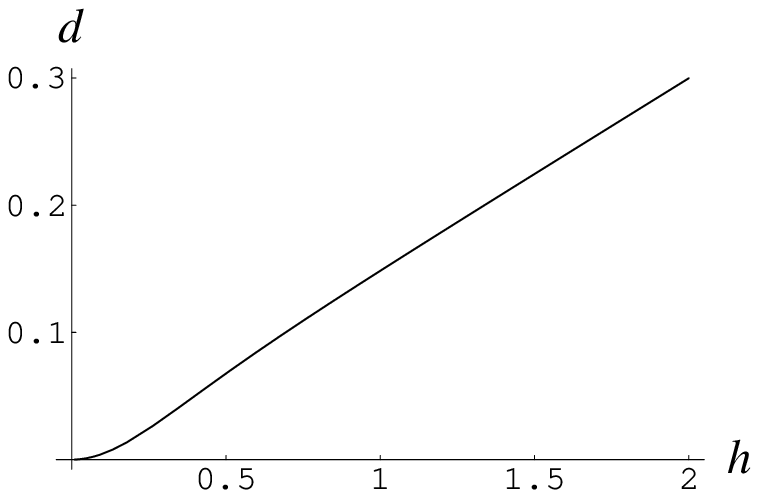,scale=0.8} \;\;\;\;\; &
\epsfig{file=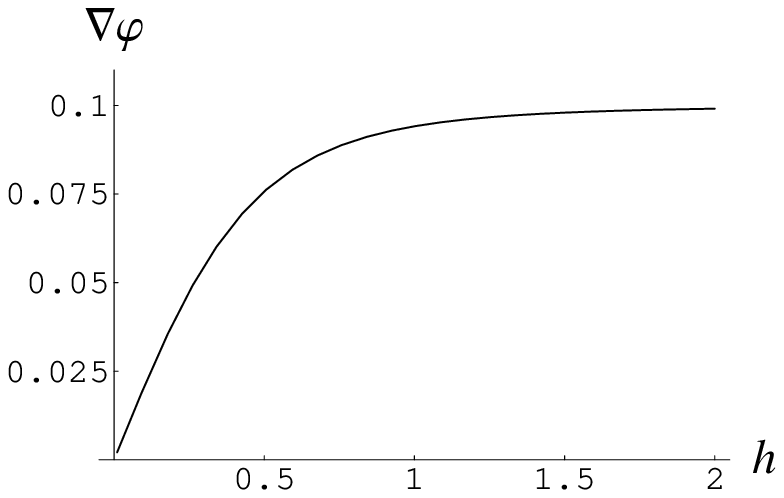,scale=0.8}
\end{tabular}
\caption{{\bf The baryon number density $d$ and the pseudo-scalar gradient $\nabla\varphi$ as 
functions of $\mu$ for fixed $h = 1$, and as functions of the magnetic field $h$ for 
fixed $\mu = 0.2$, all with $u_{KK} = 1$.}}
\label{pion_phase}
\end{center}
\end{figure}

In the absence of sources $n_{4}=0$, and all the baryon charge density comes 
from the pseudo-scalar gradient 
\be
d=\frac{3}{2} h\nabla\varphi \,.
\ee
The solution to (\ref{simple_eqn_1}) and (\ref{simple_eqn_2}) is given in this case by  
\be
a_0^V(y) &=& {\mu\over 2}\left({\cosh y_{\phantom\infty}\over\cosh y_\infty} + 1\right)\\
a_1^A(y) &=& {\mu\over 2}\, {\sinh y_{\phantom\infty}\over\cosh y_\infty} \,,
\ee
where $y_\infty \equiv y(u\rightarrow \infty)$ can be determined numerically 
in terms of $\mu$ and $h$ from the integral equation
\be
\label{integralequationpion}
y_\infty= \int_{u_{KK}}^{\infty} \frac{3h \, du}{ f^{1/2}
\sqrt{u^5 + h^2u^2 - h^2 \mu^2\, \mbox{sech}^2 y_\infty}}  \, .
\ee
The pseudo-scalar gradient is then simply
\be
\label{pion_gradient}
\nabla\varphi = {\mu\over 2}\tanh y_\infty \,.
\ee
The numerical results for $\nabla\varphi$ and $d$ as functions of $\mu$ and $h$
are presented in fig.~\ref{pion_phase}.
For $h\ll 1$, {\em i.e.} sub-string scale magnetic fields, the behavior is linear,
and the pseudo-scalar gradient is approximately given by
\be
\nabla\varphi \approx {\pi\over 2u_{KK}^{3/2}} \mu h \,.
\ee
As $h$ increases the nonlinearity of the DBI action becomes apparent.

In terms of the physical quantities we get
\be
\nabla\eta' \approx  {N_c\over 8\pi^2 f_\pi} \mu_B H 
\ee
for small magnetic fields.
This agrees with the one-flavor version of the result
(\ref{pion_gradient_son}) from \cite{Son_Stephanov}. 
The relative factor of $N_c/2$ is understood as follows.
First, the baryon charge of a quark in \cite{Son_Stephanov} is $1/N_c$
so $\mu_B^{there} = N_c\mu_B^{here}$.\footnote{Note that since $f_\pi \sim N_c$, the pseudoscalar gradient is suppressed at large $N_c$ at fixed $\mu_B^{there}$, as expected in an anomaly-mediated effect.}
Second, the CS term coupling the baryonic $U(1)_V$ gauge field to
the $\sigma_3$-component of the $SU(2)$ gauge field has a factor of 
$2$ relative to the purely abelian CS term once the boundary term at spatial infinity is omitted.

\subsection{Mixed phase}

Above some value of the chemical potential $\mu_c$ a solution with 4-brane sources
(baryons) is also possible. In this phase both the baryons and the pseudo-scalar 
gradient contribute to the baryon charge density.
We expect a phase transition to occur at this value of $\mu$ to the mixed phase.
%As in the case without a magnetic field, the critical
%chemical potential coincides with the minimal one to admit baryons.
We will compute the critical value as a function of the magnetic field $\mu_{c}(h)$, 
as well as the total baryon charge density $d(h,\mu)$ and the fraction of the 
total baryon charge carried by baryons.

Using the boundary conditions at the tip $a^V_0(y=0)=m_4/N_c$, 
$a^A_1(y=0)=0$,
the solution to the equations of motion 
(\ref{simple_eqn_1}), (\ref{simple_eqn_2}) is now given by
\be
a^V_0(y) &=& \left({m_4\over N_c}-{\mu\over 2}\right)\cosh y
+ {N_c n_4\over 3h} \sinh y +{\mu\over 2} \\[5pt]
a^A_1(y) &=& \left({m_4\over N_c} - {\mu\over 2}\right) \sinh y 
+  {N_c n_4\over 3h} \left(\cosh y - 1\right) \,.
\ee
The boundary conditions at infinity then determine the gradient and 4-brane
density implicitly in terms of $\mu$ and $h$,
\be 
\nabla\varphi &=& {\cosh y_\infty - 1\over \sinh y_\infty}\, {m_4\over N_c}\\[5pt]
\label{baryons_mixed}
N_c n_4 &=& {\frac{3}{2} h\mu + \frac{3}{2} h\left(\mu - {2m_4\over N_c}\right)
\cosh y_\infty\over \sinh y_\infty} \,,
\ee
where $y_\infty$ is the solution to to the integral equation
\be
\label{integralequation1}
y_\infty= \int_{u_{KK}}^{\infty} \frac{3h \, du}{ f^{1/2}
\sqrt{u^5\left(1+ {h^2\over u^3}\right) + \frac{9h^2}{\sinh^2 y_\infty}
\left[\left({m_{4}\over N_c}\right)^{2} 
+ \left( {\mu^2 \over 2} - {\mu m_4\over N_c}\right)(\cosh y_\infty +1) \right]}}\, .
\ee

\begin{figure}[htbp]
\begin{center}
\begin{tabular}{ccc}
\epsfig{file=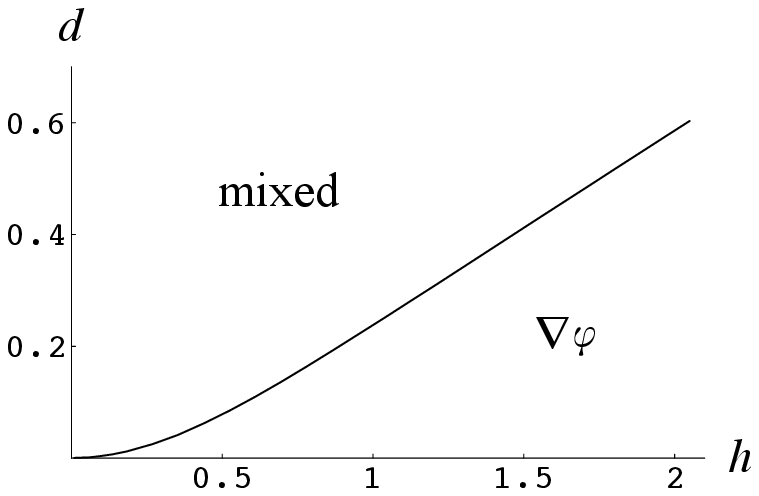,scale=0.8}
\epsfig{file=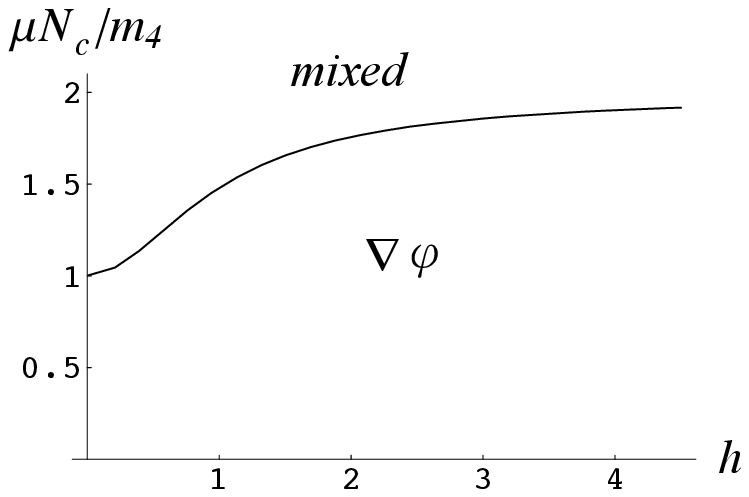,scale=0.8}
\end{tabular}
\caption{{\bf Phase diagram in the (a) canonical and (b) grand canonical ensemble.}}
\label{phase_diagram}
\end{center}
\end{figure}

The critical value of the chemical potential corresponds to the point at which
the actions of the $\nabla\varphi$ and mixed phases are equal.
But it also coincides, as it did in the zero magnetic field case, with the minimal
value of the chemical potential to create a baryon.
This can be determined numerically by setting $n_4=0$ in (\ref{baryons_mixed}).
The result is the phase diagram in the $(\mu,h)$ plane shown in fig.~\ref{phase_diagram}.
The critical chemical potential increases from its $h=0$ value $m_4/N_c$
to $2m_4/N_c$ as $h\rightarrow\infty$.
For a given $h$ there is a marginally second-order phase transition
from the $\nabla\varphi$ phase to the mixed phase at $\mu_c(h)$ that generalizes
the ordinary nuclear matter transition at $h=0$.
We also see that for a fixed total baryon charge density the pseudo-scalar gradient phase
dominates above a critical magnetic field.

\begin{figure}[htbp]
\begin{center}
\begin{tabular}{cc}
\epsfig{file=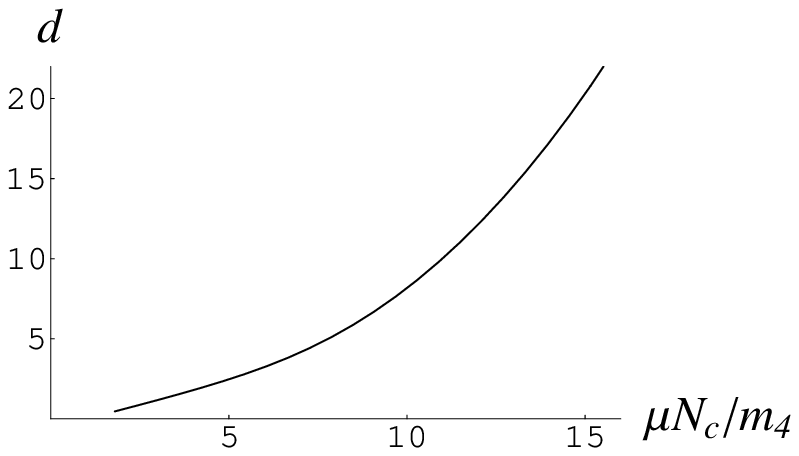,scale=0.8} \;\;\;\;\; &
\epsfig{file=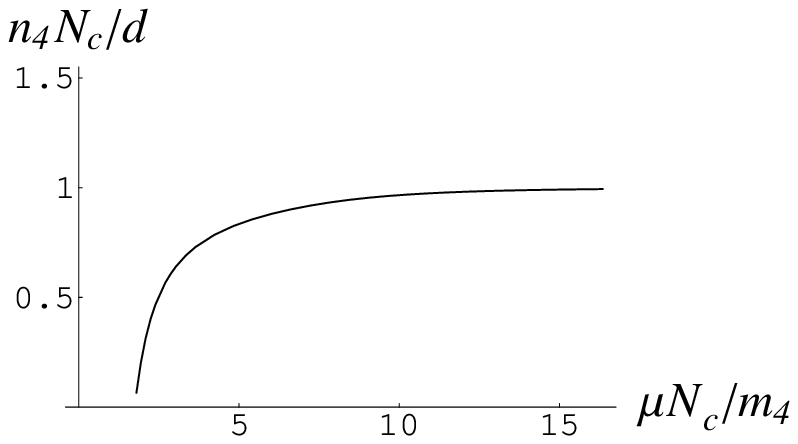,scale=0.8} \\
\epsfig{file=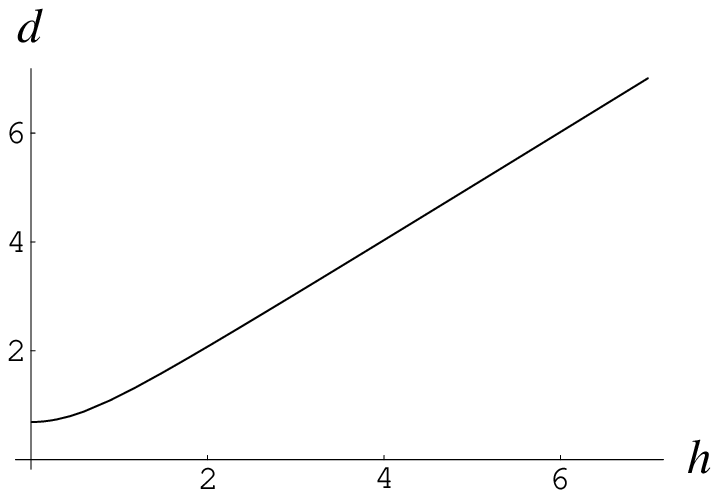,scale=0.8} \;\;\;\; &
\epsfig{file=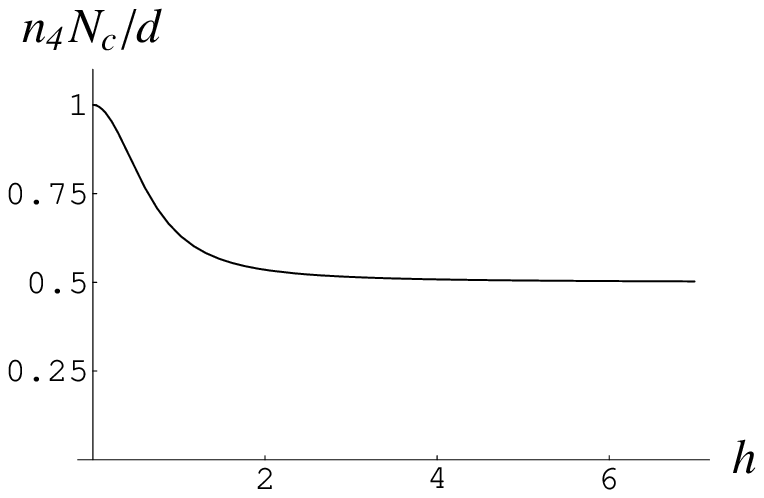,scale=0.8}
\end{tabular}
\caption{{\bf The total baryon charge density $d$ and the baryon fraction $n_4 N_c/d$, 
as functions of $\mu$ for fixed $h=1$ and $u_{KK} = 1$, and as 
functions of $h$ for fixed $\mu = 3m_4/N_c$ and $u_{KK}=1$.}}
\label{mixed_density}
\end{center}
\end{figure}

The total baryon charge density is given by
\be
\label{total_density_mixed}
d = N_c n_{4} +\frac{3}{2}h\nabla\varphi = \frac{3h}{2}\left(\mu - {m_4\over N_c}\right) \frac{\cosh y_\infty + 1}{\sinh y_\infty}
\ee
Figure \ref{mixed_density} shows the total baryon charge density
and the fraction of that charge carried by baryons, 
which are obtained by numerically computing (\ref{total_density_mixed}) and
(\ref{baryons_mixed}).
We see that the relative proportion of baryons at fixed $h$ increases with $\mu$.
In the limit of large $\mu$ the system is almost entirely baryonic nuclear matter.
On the other hand at fixed $\mu$ the proportion of baryons decreases with $h$.
For $m_4/N_c < \mu < 2m_4/N_c$ the proportion of baryons vanishes at 
the critical magnetic field $h_c(\mu)$ shown in fig.~\ref{phase_diagram}, 
where the $\nabla\varphi$ phase takes over. 

\subsection{Magnetization}

The state described by either the pseudo-scalar gradient phase or the mixed phase responds
to the external magnetic field by getting magnetized.
The {\em magnetization} $M$ can be defined in either the grand canonical
ensemble as
\be
\label{Mgrandcanonical}
M(\mu,h)= - {\frac{\partial\Omega(\mu,h)}{\partial h}\vline}_\mu \,,
\ee
where the grand potential $\Omega(\mu,h)= {S[a_0(u),a_1(u)] \vline}_{EOM}$,
or in the canonical ensemble as
\be
\label{Mcanonical}
M(d,h) = - {\frac{\partial F(d,h)}{\partial h}\vline}_d \,,
\ee
where the free energy $F = \Omega + \mu d$.
%The two definitions can be shown to be equivalent by substituting $\mu(d)$
%for $\mu$ and noting that ${\frac{\partial \Omega}{\partial \mu}\vline}_h = -d$. 
The {\em magnetic susceptibility} describes the {\em linear} response of the 
system to small magnetic fields and is defined as
\be
\label{chidef}
\chi={\frac{\partial M}{\partial h} \vline}_{h=0} \,,
\ee
in either the canonical or grand canonical ensemble.

\begin{figure}[htbp]
\begin{center}
\begin{tabular}{cc}
\epsfig{file=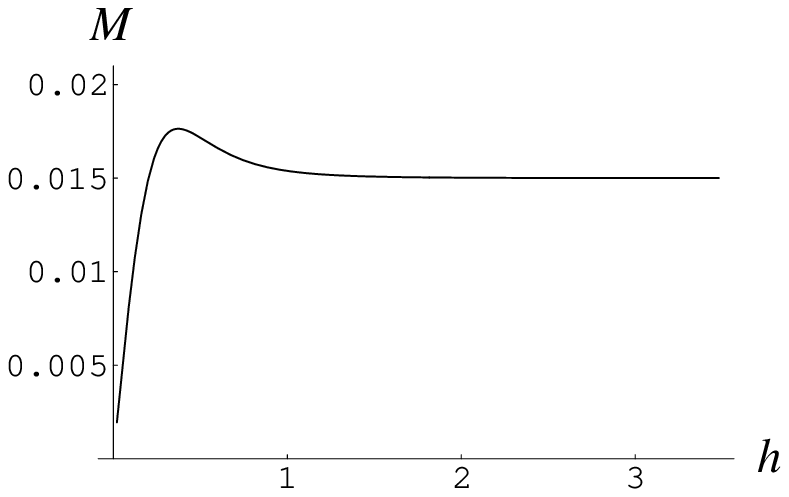,scale=0.7} \;\;\;\;\; &
\epsfig{file=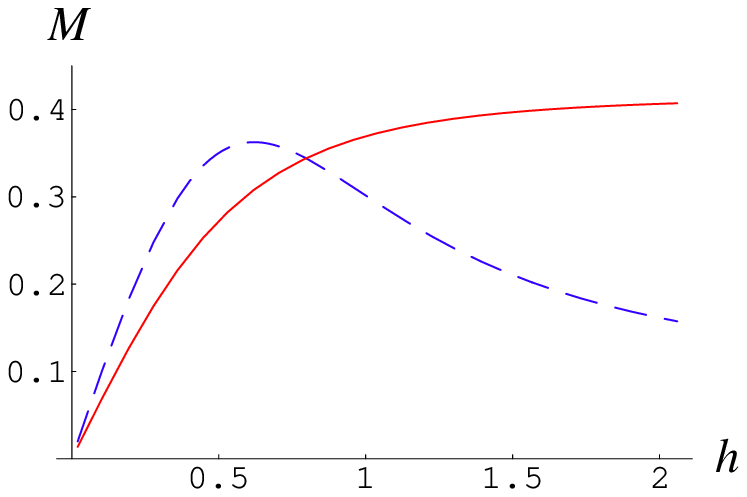,scale=0.7} \\
\end{tabular}
\caption{{\bf The magnetization $M$ (in units of ${\cal N}$) as a function of $h$ in (a) the pseudo-scalar gradient phase for fixed $\mu = 0.2$ and (b) the mixed phase for fixed $\mu = 3m_4/N_c$ (red) and fixed $d =1 $ (dashed blue), all with $u_{KK}=1$.}}
\label{M_confined}
\end{center}
\end{figure}

We would like to focus here on the matter contribution to the magnetization
and susceptibility.
The magnetic properties of the vacuum were studied in \cite{BLL2,Johnson_Kundu}.
We will therefore subtract from the quantities above the formally divergent
contribution of the vacuum, which gives a finite result that represents 
the corresponding contribution of just the matter.
%\be
%\Delta \chi =\chi(\mu)-\chi(\mu=0) \ \ \  {\rm or} \ \ \  \chi(d)-\chi({\rm vaccum})
%\ee
The numerical results for the magnetizations in the pseudo-scalar gradient and mixed phases
are presented in fig.~\ref{M_confined}.  
For small $h$ the response is linear, but for $h \sim O(1)$ the non-linear effect of the DBI action
becomes pronounced.

\begin{figure}[htbp]
\begin{center}
\begin{tabular}{cc}
\epsfig{file=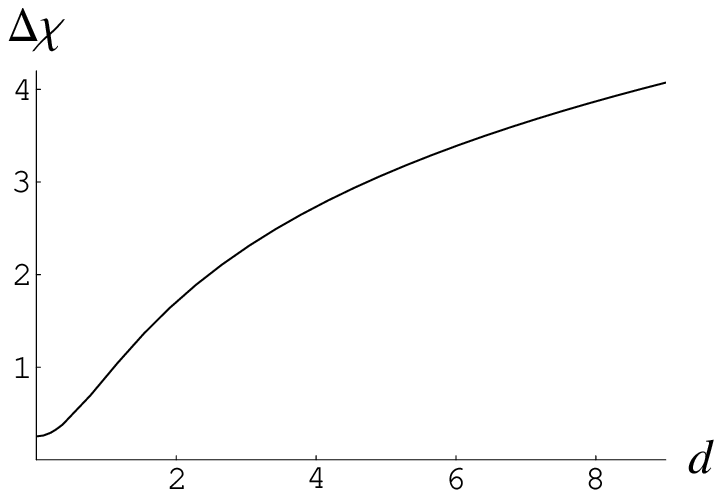,scale=0.8}\;\;\;\;\; &
\epsfig{file=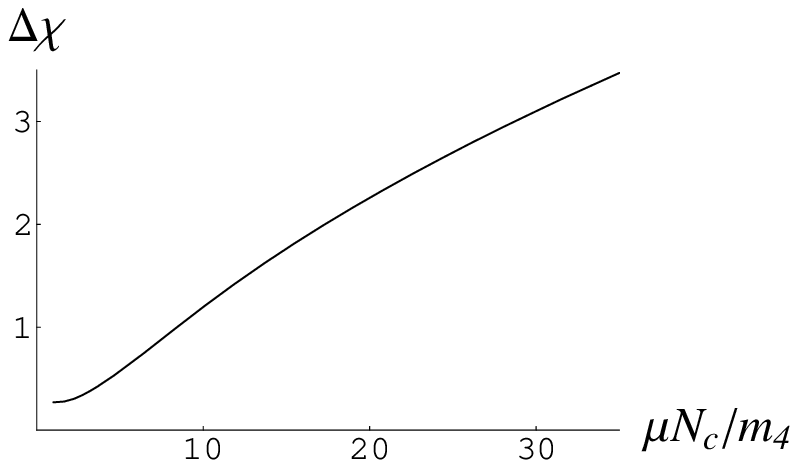,scale=0.8}\\
\end{tabular}
\caption{{\bf The magnetic susceptibility $\Delta\chi$ (divided by ${\cal N}$) of the mixed phase as a function of $d$ and $\mu$ for $u_{KK}=1$.}}
\label{chi_mixed}
\end{center}
\end{figure}

The magnetic susceptibilities can be computed in a similar way.
In the $\nabla\varphi$ phase the grand-canonical magnetic susceptibility can actually
be determined analytically to be
\be
\Delta\chi = \chi(\mu) - \chi(0) = \frac{3\pi {\cal N}\mu^2}{4 u_{KK}^{3/2}} \,.
\ee
The canonical susceptibility in this phase can be computed numerically, although we will not do this here.
In the mixed phase, the magnetic susceptibility can only be computed numerically.
Figure \ref{chi_mixed} shows the susceptibility in the mixed phase,
in both the canonical and grand canonical ensembles.
%The numerical computation of the grand-canonical magnetic susceptibility is technically more difficult,
%but can also be done, although we leave it out.
Our results show that the matter is paramagnetic in both phases.
Interestingly, the contribution to the susceptibility from the DBI term
is negative, {\em i.e.} diamagnetic, in both phases. However the contribution
of the CS term is always positive and larger.
%One can also compute the contributions to the susceptibility separately from the DBI action and from the CS action.  In both phases, for small $h$ the contribution from the DBI is always diamagnetic ($M<0$).  
%While in the pion gradient phase, as $h$ increases the DBI contribution eventually becomes paramagnetic.  However, the CS contribution is always paramagnetic and sufficiently large to overshadow the DBi term.

%%%%%%%%%%%%%%%%%%%%%%%%%%%%%%%%%%%%%%%%

\section{Magnetic properties of the deconfined phase}

The 8-brane DBI and CS actions in the deconfined background are given by
\be 
S_{DBI} &=&   {\cal{N}} \int_{u_T}^\infty du \, u^{5/2} \sqrt{\Big(1
-(a_0'(u))^2 + f(u)(a_1'(u))^2 \Big)\left(1+{h^2\over u^3}\right)} \\[5pt]
S_{CS} & = &  - {\cal N}
\int_{u_T}^\infty \left(\partial_2 a_3^V a_0^V(u) a_1^{A\prime}(u) 
-\partial_2 a_3^V a_0^{V\prime}(u) a_1^A(u)-a_{3}^{V}\partial_2 a_0^V a_1^{A\prime}+a_{3}^{V}\partial_2 a_1^A a_{0}^{V\prime}\right), \nonumber
\ee
where now $f=1-(u_T^3/u^3)$,
and where we have included both the 8-brane and anti-8-brane parts,
with $\bar{a}_0 = a_0$ and $\bar{a}_1=-a_1$.
The boundary value of the axial field is now a parameter, rather than a field, in the gauge
theory, which we set to zero, $a_1(\infty)=0$.
As before in the confined phase, we modify the Chern-Simons action by throwing away boundary terms of the form
\be
\frac{1}{2} \partial_{2}\left(a_{3}a^{}_{[1}a^{'}_{0]}\right) + \frac{1}{2} \partial_{u}\left(a_{3}\partial_{2}a_{[1}a_{0]} \right) 
\ee
to obtain the correct five-dimensional currents.

The integrated equations of motion are then given by
\be 
\label{EOMa0dec}
\frac{\sqrt{u^5+h^2 u^2}\, a_0'(u)}{\sqrt{1 - (a_0'(u))^2 + f(u)(a_1'(u))^2}}
&=& 3ha_1(u) + d\\[5pt]
\label{EOMa1dec}
\frac{\sqrt{u^5+h^2 u^2}\, f(u) a_1'(u)}{\sqrt{1 - (a_0'(u))^2 + f(u)(a_1'(u))^2}}
&=& 3ha_0(u) + j_A - \frac{3}{2}h\mu \,,
\ee
where $d$ is the baryon charge density and $j_A$ is the axial current 
density.
% so the baryon number density and axial current are related to
%the integration constants as
%\be
%d=c_0 \; , \; j_A=c_1 + h\mu \,.
%\ee
Unlike in the confined phase, we do not get an additional condition by 
extremizing the action with respect to $a_1(\infty)$.
However there is an additional condition imposed by regularity at the horizon, $a_0(u_T)=0$.
Since $f(u_T)=0$ as well, the consistency of the $a_1$ equation of motion
(\ref{EOMa1dec}) requires turning on a specific axial current density
\be
j_A = \frac{3}{2}h\mu \,.
\ee
In terms of physical quantities, the axial current density is
\be
J_A =  {N_c\over 4\pi^2}    H \mu_B \,,
\ee
which agrees precisely with the result (\ref{axial_current_son})
of \cite{axial_current}, once we account for
the different normalizations (the relative factor of $N_c/2$)
as in the pseudo-scalar gradient case in the previous section.

\begin{figure}[htbp]
\begin{center}
\begin{tabular}{ccc}
\epsfig{file=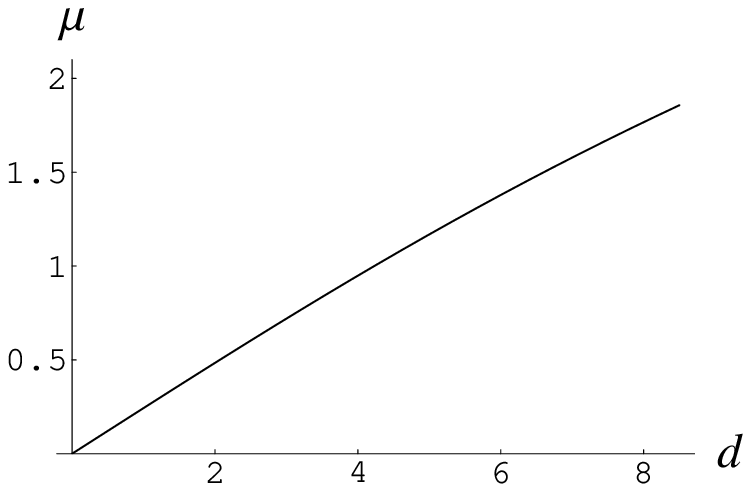,scale=0.5} \;\;\;\;\; &
\epsfig{file=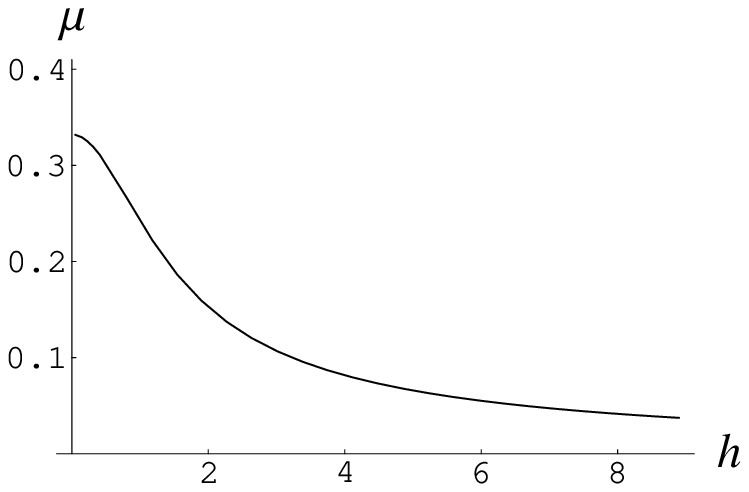,scale=0.5}  \;\;\;\;\; &
\epsfig{file=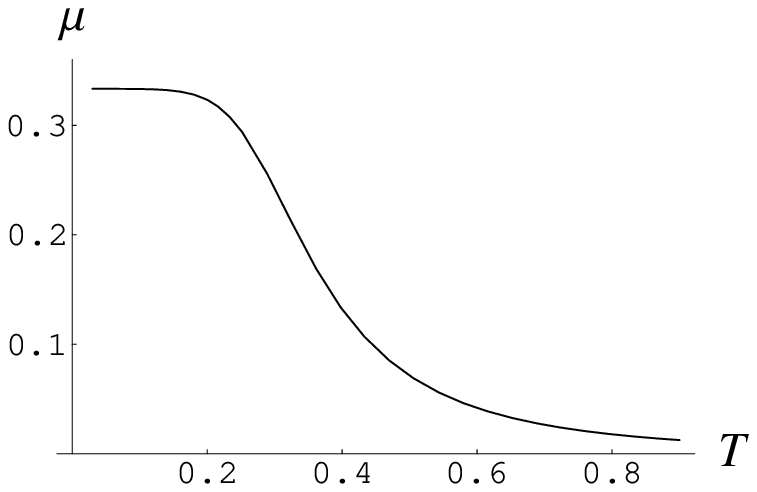,scale=0.5}
\end{tabular}
\caption{{\bf The baryon chemical potential $\mu$ as a function of (a) $d$ with fixed $h = 1$ and $T = 0.3$, (b) of $h$ with fixed $d = 1$ and $T = 0.3$, and (c) of $T$ with fixed $h=1$ and $d=1$.}}
\label{mu_parallel}
\end{center}
\end{figure}

The coupled equations of motion can be solved numerically using a shooting algorithm.  
The results for $\mu(d,h,T)$ are shown in figure \ref{mu_parallel}.
We see that at a fixed $h$, $\mu$ grows linearly with $d$ for small $d$. 
This is the expected behavior of free massless fermions in $1+1$ dimensions,
which one may think is a natural result for massless fermions in $3+1$ dimensions
in the background of a large magnetic field.
It is interesting to note however that the linear region extends to 
$d\sim h^{1.7}$, which is beyond the linear region
for free fermions that extends to $d\sim h$.
The deviation is not surprising since the fermions are not free.
We see also that, at a fixed $d$, $\mu$ decreases with $h$.
This reflects the increase of the ground state degeneracy with $h$.

\begin{figure}[htbp]
\begin{center}
\begin{tabular}{cc}
\epsfig{file=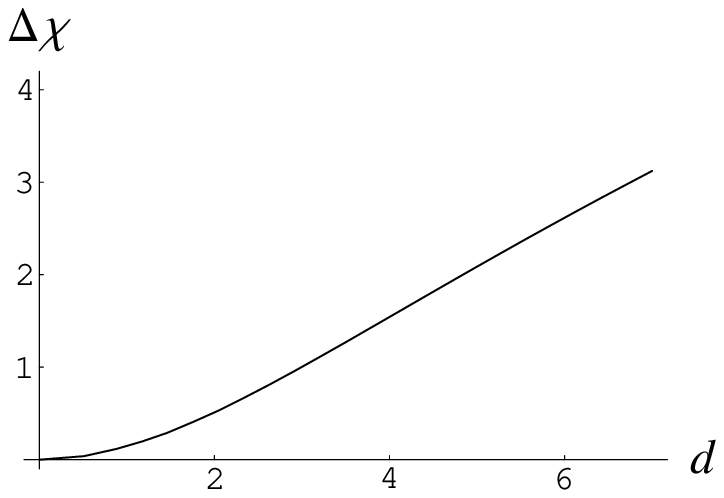,scale=0.8} \;\;\;\;\; &
\epsfig{file=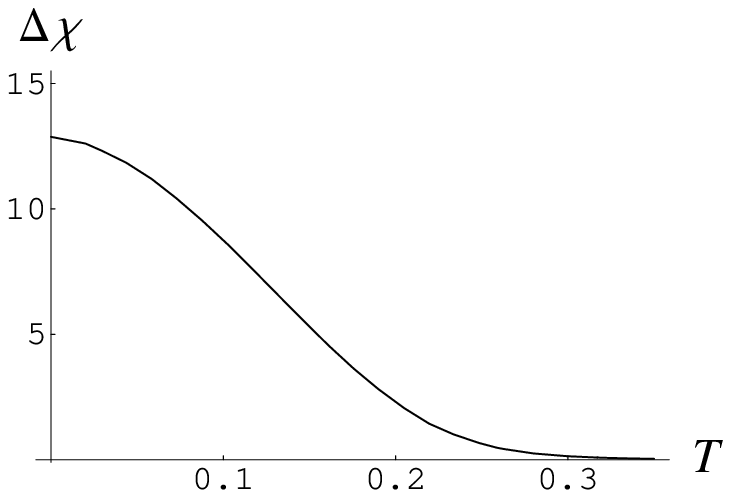,scale=0.8} \\
\end{tabular}
\caption{{\bf The magnetic susceptibility $\Delta\chi$ (divided by ${\cal N}$) of the deconfined phase as a function of (a) $d$ for fixed $T = 0.3$ and (b) as a function of $T$ for fixed $d=1$ .}}
\label{chi_parallel}
\end{center}
\end{figure}

\begin{figure}[htbp]
\begin{center}
\epsfig{file=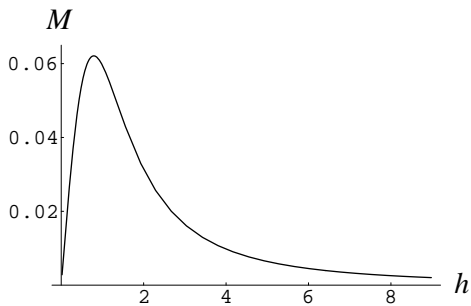,scale=0.8}
\caption{{\bf The magnetization $M$ (divided by ${\cal N}$) as a function of the magnetic field $h$ in the deconfined phase for fixed $d = 1$ and $T=0.3$.}}
\label{M_parallel}
\end{center}
\end{figure}

The magnetic response of the deconfined phase is quite similar to that of the mixed confined phase.  The magnetization and magnetic susceptibility are defined, as in the confined phase, by (\ref{Mcanonical}) and (\ref{chidef}), now with the susceptibility of the deconfined vacuum subtracted.  The numerical results for $\Delta\chi(d,T)$ and $M(h)$ computed in the canonical ensemble for fixed $d$ are shown in figures \ref{chi_parallel} and \ref{M_parallel}.
In particular, the high-temperature behavior of the susceptibility at fixed density
is $\chi \sim 1/T^9$, which deviates from the Curie law $\chi \sim 1/T$.

Finally, it is also interesting to note that the equations of motion (\ref{EOMa0dec}) and 
(\ref{EOMa1dec}) are almost symmetric under the interchange of axial and vector components.  
We considered a vector $a_0$ and an axial $a_1$, with a
chemical potential only for the vector charge $\mu = a_0(\infty)$, leading to an axial current $j_A$.  
However, if we consider instead an axial $a_0$ and a vector $a_1$, with 
an axial chemical potential $\mu_A = a_0(\infty)$, 
this would lead to a non-zero baryon number current $j_B=\frac{3}{2}h\mu_A$.

\section{Conclusions}
In this paper we have explored the properties of one-flavor holographic QCD at finite
density in a background magnetic field.  It turns out that this system has a 
rich phenomenology. In particular, in the confined phase turning on a 
magnetic field induces a gradient for the pseudo-scalar field. This gradient
carries baryon charge, and at large enough magnetic fields it
is the dominant phase. That is, if we start at zero field with some baryons, as we
increase the field those baryons will start being replaced by a gradient of the $\eta'$ field, 
eventually disappearing altogether.  In the chiral-symmetric deconfined phase 
we found that the magnetic field induces an axial current whose value is
independent of the temperature. The first property can be 
traced to the axial anomaly of fermions, and the second phenomenon can be traced
(at weak coupling) to the existence of particular fermionic zero modes 
in a magnetic field background. In the holographic dual both of these properties
are induced by the Chern-Simon term on the 8-brane, but the second 
is also due to the presence of a horizon in the spacetime geometry.

\section*{Acknowledgments}
We wish to thank D.~T.~Son for useful discussions.
This work was supported in part by the
Israel Science Foundation under grant no.~568/05.
OB gratefully acknowledges support from the Institute for Advanced Study.
OB also thanks the Institute for Nuclear Theory at the University of Washington
for its hospitality and the US Department of Energy for partial support during the 
completion of this work.

\appendix

%%%%%%%%%%%%%%%%%%%%%%%%%%%%%%%%%%%%%%%%%%%%%%%%%%%%%%%%%%%%%%%%%%%%%%%%%%%%%%%%%%%%%%%%%%%%%%%

\section{Dimensional translation table}

%\begin{table*}[htbp]
\begin{center}
\begin{tabular}{|c|c|c|}
  \hline
   quantity & dimensionless variable & physical variable \\
   \hline
  coordinates & $x_\mu,u$ & $X_\mu=Rx_\mu, U=Ru$ \\[5pt]
  gauge field & $a_\mu$ & $A_\mu = {R\over 2\pi\alpha'} a_\mu$ \\[5pt]
  magnetic field & $h$ & $H={1\over 2\pi\alpha'} h$ \\[5pt]
  baryon chemical potential & $\mu$ & $\mu_B={R\over 2\pi\alpha'}\mu$ \\[5pt]
  baryon charge density & $d$ & $D={2\pi\alpha'{\cal N}\over R} d$ \\[5pt]
  axial current density & $ j_A$ & $J_A={2\pi\alpha'{\cal N}\over R} j_A$ \\[5pt]
  pseudo-scalar field & $\varphi$ & $\eta' = {R^2 f_\pi\over 2\pi\alpha'} \varphi$ \\[5pt]
  wrapped 4-brane mass & $m_4$ & $M_4 = {R\over 2\pi\alpha'} m_4$  \\[5pt]
  wrapped 4-brane density & $n_4$ & $N_4 = {2\pi\alpha'{\cal N}\over R} n_4$\\[5pt]
  \hline
\end{tabular}
\end{center}
%\end{table*}

\end{document}